\documentclass[twocolumn,prl,floatfix]{revtex4-1}
\usepackage{color}
\usepackage{graphicx}
\usepackage{amsmath}
\usepackage{amssymb}
\usepackage{mathtools}
\usepackage{wasysym}

\newcommand{\Tr}{\mathop{\mathrm{Tr}}}

\newcommand{\Dec}{\mathcal{D}}

\newcommand{\rhoss}{\rho_{\rm ss}}

\newcommand{\rhossp}{\rhoss^{\rm p}}

\newcommand{\nmode}{n_{\rm m}}

\newcommand{\be}{\begin{equation}}
\newcommand{\ee}{\end{equation}}
\newcommand{\nn}{\nonumber}

\newcommand{\Phic}{\Phi_c}
\newcommand{\Phis}{\Phi_s}

\begin{document}

\title{Spin Correlations as a Probe of Quantum Synchronization in Trapped Ion Phonon-Lasers}

\author{Michael R. Hush$^{1,2}$}
\author{Weibin Li$^1$}
\author{Sam Genway$^1$}
\author{Igor Lesanovsky$^1$}
\author{Andrew D. Armour$^1$}

\affiliation{$^1$School of Physics and Astronomy, University of Nottingham, Nottingham NG7 2RD, United Kingdom}
\affiliation{$^2$School of Information Technology and Electrical Engineering, University of New South Wales at the Australian Defence Force Academy, Canberra, Australia}


\date{\today}

\begin{abstract}
	We investigate quantum synchronization theoretically in a system consisting of two cold ions in microtraps. The ions' motion is damped by a standing-wave laser whilst also being driven by a blue-detuned laser which results in self-oscillation. Working in a non-classical regime, where these oscillations contain only a few phonons and have a sub-Poissonian number variance, we explore how synchronization occurs when the two ions are weakly coupled using a probability distribution for the relative phase. We show that strong correlations arise between the spin and vibrational degrees of freedom within each ion and find that when two ions synchronize their spin degrees of freedom in turn become correlated. This allows one to indirectly infer the presence of synchronization by measuring the ions' internal state.
\end{abstract}

\maketitle

\emph{Introduction.} Two macroscopic self-oscillators synchronize when  their {\it relative} phase locks to a fixed value\,\cite{Pikovsky2003}. Important studies of synchronization effects were carried out using lasers\,\cite{Roy:1994,*DeShazer:2001, *Heil:2001}, with arrays of Josephson junctions \cite{Wiesenfeld:1996, *Acebron:2005, *Vinokur:2008} and over the last few years much attention has been devoted to exploring synchronization in micromechanical oscillators~\cite{Agrawal:2013,*Roukes:2014, *Giesler:2014}. Recently, theoretical work has begun to explore synchronization in the quantum regime~\cite{Lee2013,Lee2014,Walter:2014,Walter:2014a,Choi:2014,Mari2013,Xu:2014,Ludwig:2013,Manzano2013,Mendoza2013}: the formation of a relative phase preference between two (or more) weakly coupled quantum oscillators operating in a regime far from the classical correspondence limit. Differences between classical and quantum predictions for the synchronization of van der Pol oscillators have been identified in the case where the oscillators are only weakly excited\,\cite{Lee2013}. Nevertheless, many important questions about quantum synchronization remain open, such as how it should be quantified and how it can best be probed experimentally.

Cold ions in microtraps provide a natural platform for exploring synchronization in the quantum regime\,\cite{Lee2013}. The generation of self-oscillations in the motional state of ions, phonon-lasing, has already been observed\,\cite{Vahala2009,*Knunz:2010,*Xie:2013}. Furthermore, precise control of trapping potentials of the individual ions can now be achieved with microtraps\,\cite{brown_2011,*harlander_2011} allowing the vibrational frequencies of individual ions and the coupling between different ions to be tuned. Here, we investigate synchronization in two trapped ion phonon-lasers which are pumped in a similar way to that demonstrated in recent experiments\,\cite{Vahala2009,*Knunz:2010}.

We identify a parameter regime where phonon-lasing of an individual ion occurs with just a few quanta leading to a non-classical state of the phonons and investigate the emergence of synchronization in this regime when a weak inter-ion coupling is introduced (weak as it is the slowest time scale in the system). Our model includes two of the electronic levels of the ions used in the pumping process (which we refer to as `spin'), allowing us to uncover strong correlations which arise between the electronic and vibrational degrees of freedom of the individual ions. We study the degree of synchronization as the strength and detuning of the pumping lasers are varied by calculating the probability distribution for the relative phase of the ion's phonons. Lastly we show that synchronization between the ion's vibrational degrees of freedom can lead to correlations between the `spins' of the two ions. Indeed, observation of spin-correlations form a sufficient and convenient method of inferring synchronization between two phonon-lasers.

\begin{figure}[t!!]
\includegraphics[width=\columnwidth]{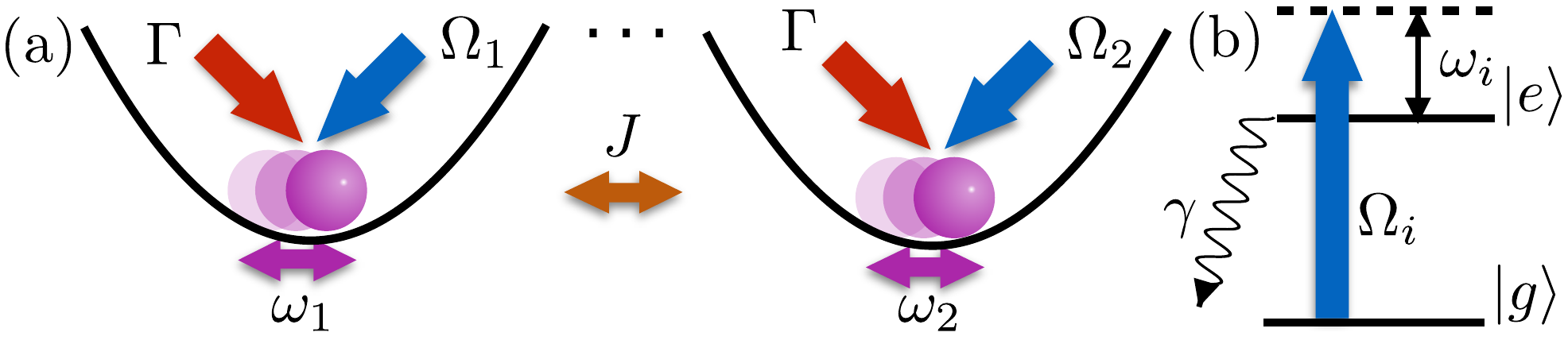}
\caption{(Color online) (a) Trapped ion setup. Each ion is damped at a rate $\Gamma$ by a standing-wave laser and driven by a blue-detuned laser of strength $\Omega_{j=1,2}$. The phonons have a dipole interaction of strength $J$ and the trap frequencies are $\omega_1$ and $\omega_2$. (b) Internal electronic states of each ion. The `spin' states are pumped by a laser blue-detuned by frequency $\omega_{j=1,2}$ and undergo spontaneous emission at a rate $\gamma$. The  damping is achieved using a red-detuned drive on a different electronic transition (not shown) and is eliminated adiabatically.}
\label{fig:pg1}
\end{figure}

\emph{Trapped Ion Setup.} A sketch of the system we study is shown in Fig.~\ref{fig:pg1}. Each ion is in a microtrap\,\cite{brown_2011,*harlander_2011} with frequency $\omega_{j=1,2}$. The quantized vibrational degrees of freedom (phonons) are linearly damped at a rate $\Gamma$, which can be realised by laser cooling techniques~\cite{Cirac:1992,*Marzoli:1994,Genway2014}. Each ion's spin (internal) degree of freedom is driven by standing wave lasers with Rabi frequencies $\tilde{\Omega}_{j=1,2}$, which are set to be resonant with the first blue sideband transition. The two ions interact weakly via a dipole interaction which leads to a linear coupling of their phonons with strength $J$ \cite{brown_2011,*harlander_2011}. In the rotating wave approximation, the dynamics of the ions is governed by the master equation
\begin{align}
\dot{\rho} = -i[H,\rho] + \sum_{j=1,2} & \Big\{ \frac{\gamma}{2} \int_{-1}^1 dz W(z) \Dec[e^{i\eta {q_j}z} \sigma^-_j] \rho \nn \\
& +  \Gamma \Dec[a_j] \rho  \Big\} \label{eqn:exactme}
\end{align}
with: 
\begin{align}
H = & \sum_{j=1,2} \frac{1}{2}\{ \omega_j (  2 a^\dag_j a_j - \sigma^z_j ) + \tilde{\Omega}_j\sigma_j^x \sin (\eta q_j) \} \nn \\
  &+ J q_1 q_2, \nn \\
W(z) =& \frac{3}{4}(1 + z^2), \nn \\
\Dec[L](\rho) = & L\rho L^\dag - \frac{1}{2}(L^\dag L \rho + \rho L^\dag L). \nn
\end{align}
Where $a_j$ are the annihilation operators for the phonons, $\sigma_j^{\alpha=x,y,z}$ are the Pauli operators for the spins, $q_j = {a}^\dag_j + {a}_j$ is the position operator, $\eta$ is the Lamb-Dicke parameter and $W(z)$ is the angular distribution for spontaneous emission. We have adjusted the laser such that the detuning with respect to the spin is $-\omega_j$.

We simplify the master equation by assuming that the system is in the Lamb-Dicke regime ($\eta\ll 1$), and retain only terms up to and including first-order in $\eta$.  We also assume that the trapping potentials are tight, $\omega_j\gg \gamma,\Gamma,\Delta, \eta\tilde{\Omega}_j$. This allows us to neglect terms that are rotating at or above the mean frequency $\omega=(\omega_1+\omega_2)/2$ after we move to a frame rotating at the frequency $\omega$. This leads to the following simplified master equation
\begin{align}
\dot{\rho} \approx & -i[H_e,\rho]  + \sum_{j=1,2} \left\{ \gamma \Dec[\sigma^-_j] (\rho) + \Gamma \Dec[a_j](\rho)\right\}, \label{eqn:mainmast}
\end{align}
with:
\begin{align}
H_e \approx & \sum_{j=1,2} \frac{1}{4}\{ (-1)^{j} \Delta (2a_j^\dag a_j - \sigma^z_j ) \nn \\
& + 2\Omega_j (a^\dag_j \sigma^+_j + a_j\sigma^-_j) \}  + J (a_2^\dag a_1 +a_1^\dag a_2 ). \nn
\end{align}
Where $\sigma^{\pm}_j = (\sigma^x_j \pm i\sigma_j^y)/2$, $\Omega_j = \eta \tilde{\Omega}_j$, and $\Delta = \omega_2 - \omega_1$. Hence the spin-photon coupling is described by the anti-Jaynes-Cummings Hamiltonian $H_e$. After simplification, we see that using a standing wave configuration (as opposed to a running wave) means that the leading contribution from the spin-phonon coupling expansion is linear in $\eta$, while higher order terms can be neglected safely. The numerical results described below were all obtained using the steady-state solution to Eq.~(\ref{eqn:mainmast}), $\rhoss$\,\cite{Johansson:2012, *Johansson:2013}.

\begin{figure}[t!]
\includegraphics[width=\columnwidth]{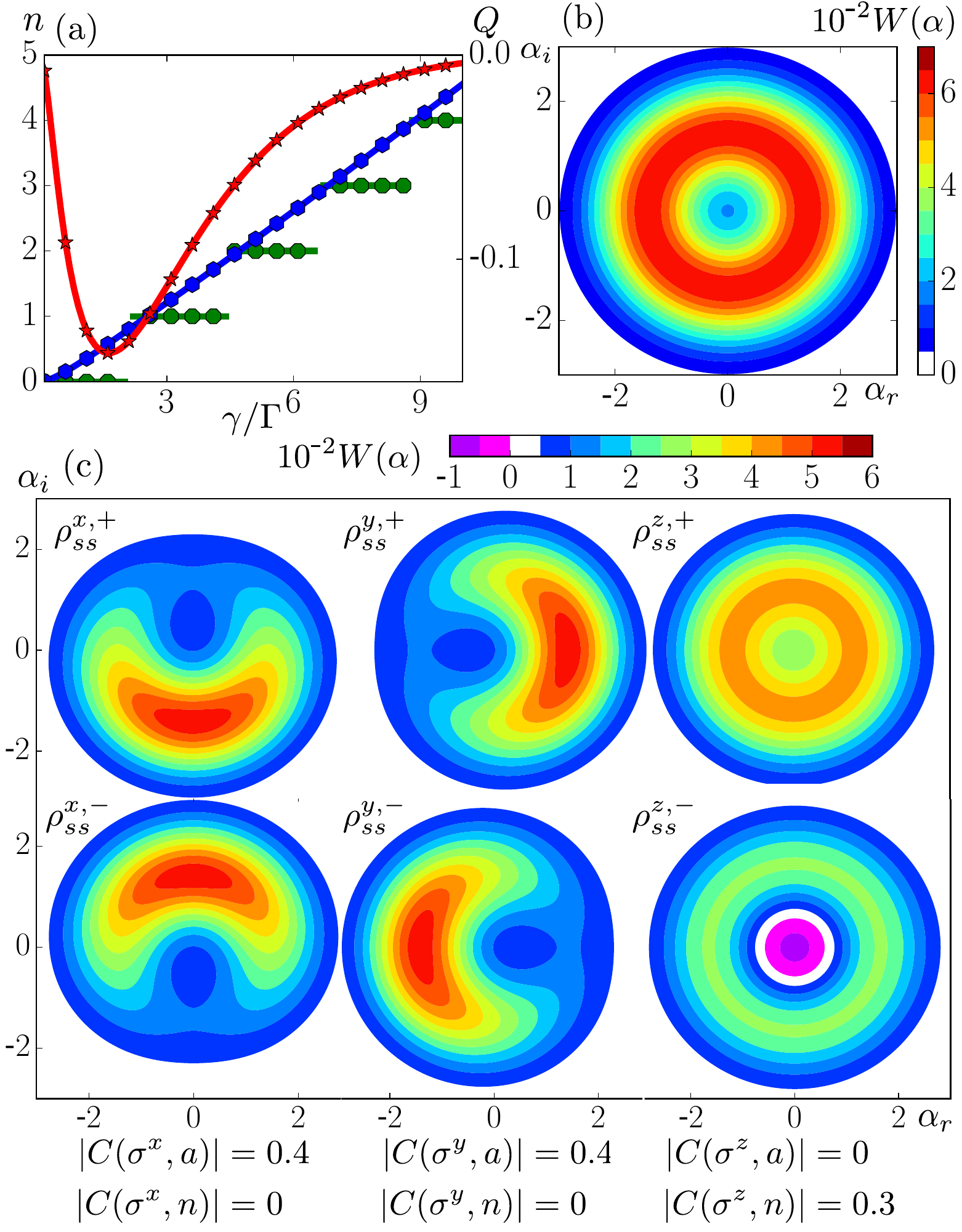}
\caption{(Color online) (a) Average phonon number $\langle n\rangle$ (blue hexagons) and most likely phonon number $\nmode$ (green octogons) of the phonon distribution plotted against damping strength $\gamma/\Gamma$, we also plot the Mandel Q parameter (red stars) which becomes negative around the lasing transition. (b) Wigner distribution of the phonons for $\Gamma/\gamma = 1/3$. (c) Wigner distribution of the phonons after projection onto different spin eigenstates $\rhoss^{j,\pm}$ ($\Gamma/\gamma = 1/3$). $\sigma^{x/y}$ operators are correlated with phase and $\sigma^z$ operators are correlated with number, which is confirmed by calculating the correlations, given beneath their corresponding Wigner distributions. Here $\Omega/\gamma = 1$, and $\Delta=0$ throughout.}
\label{fig:pg2}
\end{figure}

\emph{Individual Ions.} A prerequisite for synchronization is that each individual ion undergoes self-oscillations in their motion, so-called phonon lasing \cite{Vahala2009,*Knunz:2010,*Xie:2013}. When the phonons are driven sufficiently strongly to overcome the damping, $\Omega^2> \gamma \Gamma$, the mean-field equations of motion show a limit-cycle solution with  $\langle n \rangle = \gamma /2\Gamma - \gamma^2 /2\Omega^2$ where $n = a^\dag a$ \cite{Gardiner:2004}. We have confirmed these mean-field predictions numerically by finding the steady state of an ion for fixed driving $\Omega$ and decreasing damping rate $\Gamma$ ($\Delta = 0 $ and $J=0$). The onset of phonon-lasing can be seen using the average phonon number $\langle n\rangle$ and the phonon number which is most likely to be observed $\nmode$. In Fig.~\ref{fig:pg2}(a) we see both these parameters get larger as the damping is decreased. The onset of lasing is also visible in the Mandel-Q parameter, $Q = (\langle n^2 \rangle - \langle n \rangle^2) /\langle n \rangle - 1$. Moreover we see sub-Poissonian statistics around the lasing transition; this is because we are using a single two-level system for the pump \cite{Dubin:2010}. In the following we investigate synchronization in a quantum regime where $\Omega/\gamma = 1$ and $\Gamma/\gamma = 1/3$;  here $\langle n\rangle= 1.2$, $\nmode = 1$ and $Q=-0.1$. The steady-state Wigner distribution \cite{Gardiner:2004} for these parameters is shown in Fig.~\ref{fig:pg2}(b). It has a `doughnut' shape, as the state has a non-zero average amplitude, but no phase preference.

Before investigating synchronization in coupled ions, we examine the correlations that build up between the spin and phonon degrees of freedom in an individual ion due to their strong coupling. We later exploit these correlations to show how the presence of synchronization between the ions' phonons can be inferred through measurements of their spins. The spin-phonon correlations become apparent in Fig.~\ref{fig:pg2}(c), where we plot the Wigner distribution of the density matrix for an individual ion after projection with one of Pauli-operator eigenstates. Specifically, we apply $P^{\alpha,\pm}$, where $\sigma^{\alpha} P^{\alpha,\pm}= \pm P^{\alpha,\pm}$ and $(P^{\alpha,\pm})^2 = P^{\alpha,\pm}$, to the steady state of the system: $\rhoss^{\alpha,\pm} = \Tr_s[P^{\alpha,\pm} \rhoss]$. We see that the projections onto the eigenstates of $\sigma^z$ are correlated with phonon number, but not phase. Interestingly, we see some negativity in the Wigner distribution after projection with $P^{z,-}$, which provides further evidence that we are in the quantum regime. On the other hand, projections onto the eigenstates of, $\sigma^x$ and $\sigma^y$ are correlated with the phase of the phonons, but not the number. We confirm these observations by studying the correlators $C(\sigma^{\alpha},a)$ and $C(\sigma^{\alpha},n)$ which relate to the spin-phonon phase and number correlations respectively (shown in Fig~{\ref{fig:pg2}}), with the correlation between two operators $X$ and $Y$ defined as $C(X,Y) = \langle X Y \rangle - \langle X \rangle \langle Y \rangle$.

\emph{Coupled Ions.} We now consider how synchronization arises for two weakly coupled ions. Classically synchronization originates from the development of stable fixed points in the equation of motion for the relative phase. For a quantum system, our intuition suggests that there exists some relative-phase distribution which is not flat when the ions are synchronized. A candidate phase distribution is the Wigner distribution after integrating over the radial and total phase coordinates \cite{Lee2013}. But, in general, a distribution based on a quasi-probability distributions is not unique as other representations could be used  \cite{Husimi:1940,*Sudarshan:1963,*Drummond:1980,*Vaccaro:1995,*Vaccaro:1990,*Moya-Cessa:2003,*Hush:2010}, which would give different results. We circumvent this ambiguity by directly calculating a relative phase distribution from the density matrix using phase states \cite{Barak:2005,*QOp:2007,*Pegg:1988,*Shapiro:1991}:
\begin{align}
P(\phi) & = \iint_0^{2\pi} d\phi_1 d\phi_2 \; \delta(\phi_1 - \phi_2 - \phi) \langle \phi_1, \phi_2| \rhossp | \phi_1,\phi_2 \rangle \nn \\
& = \sum_{\mathclap{n,m=0}}^{\infty} \; \frac{e^{i(m-n)\phi}}{2\pi} \; \sum_{\mathclap{d =\max(n,m)}}^{\infty} \; \langle n,d-n| \rhossp |m,d-m \rangle \label{eqn:Pphidefn}
\end{align}
where $|\phi _j\rangle = \sum_{n=0}^{\infty} e^{i\phi_j n}/\sqrt{2\pi} |n\rangle $ and $ \rhossp = \Tr_s[ \rhoss ]$ is the steady-state density matrix after tracing over the spins. $P(\phi)$ is positive and normalized.

\begin{figure}[t!]
\includegraphics[width=\columnwidth]{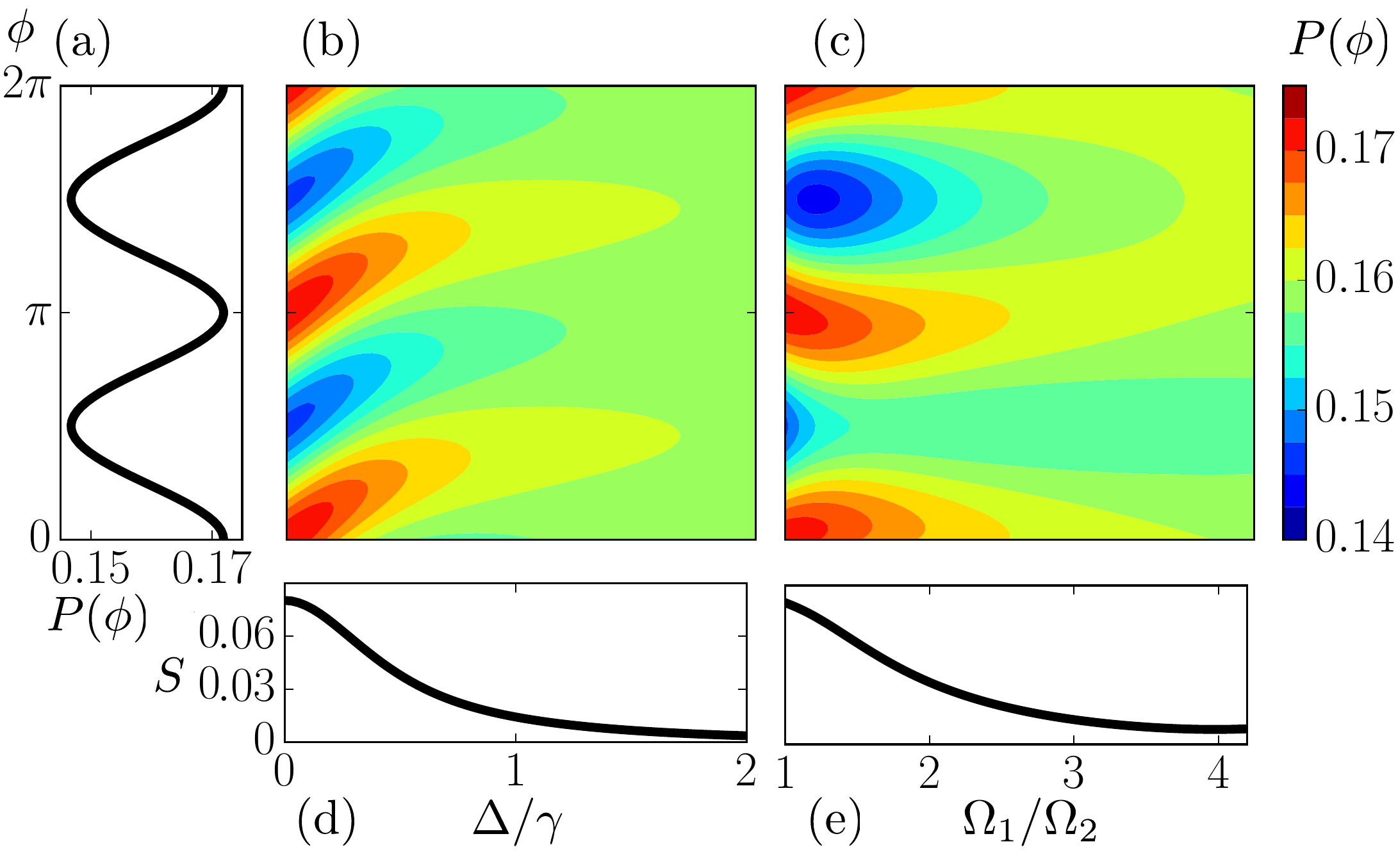}
\caption{(Color online) Phase distribution $P(\phi)$: (a) when the system is perfectly symmetric ($\Omega_1/\Omega_2 = 1$, $\Delta = 0$); (b) while the oscillators are detuned $\Delta/\gamma$ ($\Omega_1/\Omega_2 = 1$); and (c) for different relative pumping strengths $\Omega_1/\Omega_2$ ($\Delta = 0$). In (d) and (e) the synchronization measure $S$ (black solid) is plotted against the same parameters as (b) and (c) respectively. Here $\Omega_2/\gamma = 1$, $\Gamma/\gamma = 1/3$ and $J/\gamma=1/10$ throughout.}
\label{fig:pg3}
\end{figure}

We look for a signature of synchronization by calculating the relative phase distribution $P(\phi)$ from the steady state solution to the master equation when ions are in the lasing regime and weakly coupled: $J/\gamma=1/10$. We plot $P(\phi)$ in Fig.~\ref{fig:pg3} with different values for $\Delta$ and $\Omega_1/\Omega_2$:

\noindent
Firstly, we consider the symmetric case ($\Delta = 0$ and $\Omega_1 /\Omega_2= 1$) for which $P(\phi)$ is shown in Fig~\ref{fig:pg3}(a). The ions show signatures of synchronization as $P(\phi)$ is not flat, in fact the distribution is bimodal and $\pi$-periodic. This bimodal feature is analogous to the bistability typically seen in synchronized classical systems with inertial coupling \cite{Kuznetsov:2009}.

Next we consider detuning of the oscillators frequencies (see Fig~\ref{fig:pg3}(b)). Here we see the maximum height of $P(\phi)$ gets smaller as it approaches a flat (unsynchronized) distribution. The $P(\phi)$ distribution stays exactly $\pi$-periodic and bimodal during this process, although the phase of the distribution does shift. Loss of synchronization due to detuning is also seen in classical systems \cite{Kuznetsov:2009}; but it is typically a sharp transition (in the absence of thermal noise), while we see in the quantum case it is smooth \cite{Walter:2014a,Walter:2014}.

Lastly, when one of the ions is pumped more strongly than the other such that $\Omega_1>\Omega_2$ (while $\Delta = 0$), we find that $P(\phi)$ changes continuously from being bimodal to unimodal (see Fig~\ref{fig:pg3}(c)). Such transitions between monostable and bistable synchronized states have also been observed in classical systems with unequal driving \cite{Ivanchenko:2004}.

In order to quantify of synchronization for a given $\rhoss$, we propose a simple measure based on the relative phase distribution:
\be
S = 2\pi \max[P(\phi)] - 1.
\ee
$S$ is in essence the peak height of $P(\phi)$ above a flat distribution. It is a useful measure for synchronization as it is non-zero if and only if $P(\phi)$ is not flat, which we regard as the signature for the presence of synchronization. We plot $S$ in Fig.~\ref{fig:pg3}(d) and~(e) to summarize the strength of synchronization in the system. In Fig.~\ref{fig:pg3}(d) we see $S$ goes to zero as the two oscillators are increasingly detuned and synchronization vanishes.


\emph{Synchronization and Spin Correlations.} We have shown that synchronization is present in the motional state of our ions using the distribution $P(\phi)$ and the related measure $S$. Sophisticated experimental techniques have been developed to perform full state-tomography of an ion's phonons which would give access to $\rhossp$ \cite{Leibfried:2003}, from which $P(\phi)$ and $S$ could be calculated. Nevertheless, the spin-phonon locking seen in Fig.~\ref{fig:pg2}(c) suggests we may be able to infer the presence of synchronization indirectly through measurements \cite{Leibfried:2003} of the spin degrees of freedom alone.

The spin-phonon locking occurs faster than the phonon-phonon synchronization as the inter-ion coupling provides the longest time-scale in the system. Thus it is reasonable to make a measurement of each ion's spin and use our \emph{a priori} knowledge of the spin-phonon locking to infer the phase of the oscillators. We focus on the spin-spin correlations in particular, which we split into two types: `number correlations' (e.g. $C(\sigma_z^1,\sigma_z^2$) and `phase correlations' (e.g. $C(\sigma_x^1,\sigma_x^2)$  and $ C(\sigma_x^1,\sigma_y^2)$).

The phase correlations can be related semi-classically to statistical moments of $P(\phi)$, which in turn give us information about the ion's state of synchronization. Solving the mean field dynamics for $\langle \sigma^-_j \rangle$ in terms of $\langle a_j \rangle$ gives $\langle \sigma^-_j \rangle \propto -i e^{-i\phi_j}$ with $\phi_j=\arg[\langle a_j \rangle]$, in agreement with Fig.~\ref{fig:pg2}(c). Although, strictly speaking all correlations are zero in a mean-field calculation, we can use this mean-field equality as an ansatz to describe how the spin and phonon \emph{operators} are related in the steady state. We assume $\sigma^-_j\propto -i{\rm e}^{-i\phi_j} $ and then calculate the expectation values by taking an average over the corresponding steady-state phase distributions (remembering that there is no preferred total phase so its distribution is always flat).  This then leads us to the approximate relationships between the expectation values and moments of the relative phase distribution: $C(\sigma_x^1,\sigma_x^2) \propto \Phic \equiv \int d\phi \cos \phi P(\phi)$ and $C(\sigma_x^1,\sigma_y^2) \propto \Phis \equiv -\int d\phi \sin \phi  P(\phi)$. When $P(\phi)$ is flat $\Phi_{c,s} = 0$, this means $\Phi_{c,s} \ne 0$ is a sufficient (but not necessary) condition for the ions to be synchronized. Consequently, we expect that measurements of spin correlations can be used to infer the presence of synchronization.

\begin{figure}[t!]
\includegraphics[width=\columnwidth]{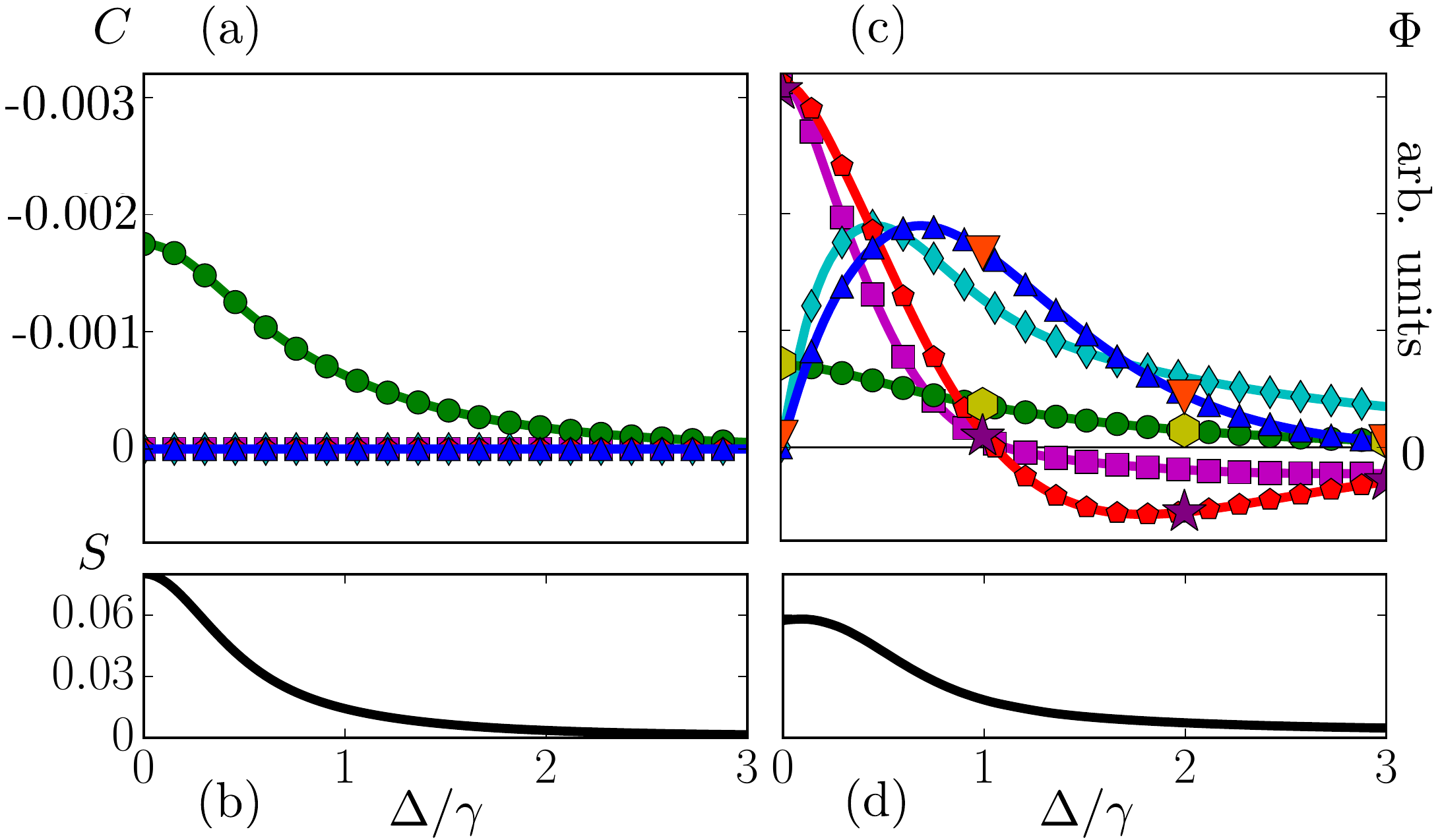}
\caption{(Color online) (a) and (c) $C(\sigma_1^z,\sigma_2^z)$ (green circles), $C(\sigma_1^x,\sigma_2^x)$ (blue up triangles), $C(\sigma_1^x,\sigma_2^y)$ (red pentagons), $\Phic$ (cyan diamonds) and $\Phis$ (magenta squares)  are plotted against $\Delta$. In (a) the driving is balanced ($\Omega_1/\Omega_2=1$) while in (c) $\Omega_1$ is stronger ($\Omega_1/\gamma= 5/4$ and $\Omega_2/\gamma = 1$). (b) and (d) $S$ (black solid) is plotted for comparison with the same parameters as (a) and (c) respectively. Here $\Omega_1/\gamma = 1$, $\Gamma/\gamma = 1/3$ and $J/\gamma = 1/10$. In (c), the red down triangles, purple stars and yellow hexagons come from calculations of $C(\sigma_1^x,\sigma_2^x)$, $C(\sigma_1^x,\sigma_2^y)$ and $C(\sigma_1^z,\sigma_2^z)$, respectively, using a version of Eq.\ (\ref{eqn:exactme})  in which we neglect only terms $O(\eta^4)$ and higher. They agree well with the results obtained using Eq.\ (\ref{eqn:mainmast}).}
\label{fig:pg4}
\end{figure}

We investigate the connection between spin correlations and synchronization in Fig.~\ref{fig:pg4}(a), where we plot $C(\sigma_1^z,\sigma_2^z)$, $C(\sigma_1^x,\sigma_2^x)$, $C(\sigma_1^x,\sigma_2^y)$, $\Phic$ and $\Phis$, as a function of the detuning $\Delta$ for equal driving strength ($\Omega_1/\Omega_2=1$). We plot $S$ for the same parameters in Fig.~\ref{fig:pg4}(b) to measure the synchronization strength. The number correlation $C(\sigma_1^z,\sigma_2^z)$ is initially negative and approaches zero as the detuning is increased. This indicates that the phonon-numbers of the oscillators are correlated at small $\Delta$, but does not directly indicate a phase relationship. Note that $C(\sigma_1^x,\sigma_2^x)$ and $C(\sigma_1^x,\sigma_2^y)$ match $ \Phic$ and $\Phis$, respectively. However, these quantities are all zero. This is an example where $\Phi_{c,s} = 0$ which means we do not have sufficient information to conclude whether the ions are synchronized or not. Indeed, $S$ is non-zero in this case. This occurs because $P(\phi)$ is $\pi$-periodic for equal driving, even as the lasers are detuned (see Fig.~\ref{fig:pg3}(b)); any phase distribution that is $\pi$-periodic will give $\Phi_{c,s}=0$ and hence $C(\sigma_1^x,\sigma_2^x)=C(\sigma_1^x,\sigma_2^y)=0$. Normally one would consider a different statistical moment of the probability distribution to circumvent this issue e.g. if one had two random variables that were uncorrelated in their averages $C(X,Y)=0$ one could determine $C(X^2,Y^2)$ and possibly find correlations in their variances. Unfortunately the Pauli operator algebra makes such an approach impossible e.g., $(\sigma^\alpha_j)^2 = 1$.

This problem can be overcome by introducing a slight asymmetry in the driving strengths, which breaks the $\pi$-periodic nature of $P(\phi)$ (see Fig.~\ref{fig:pg3}(c)). In Fig.~\ref{fig:pg4}(c) we plot the same correlations and moments as Fig.~\ref{fig:pg4}(a) against detuning, but this time with unequal driving $\Omega_1/\Omega_2 = 5/4$. The spin correlations are now all present, with the phase correlations being much stronger than the number correlations. Even though we are in a quantum regime, we see $C(\sigma_x^1,\sigma_x^2)$ and $C(\sigma_x^1,\sigma_y^2)$ have behavior that follows that of the semi-classical estimates $\Phic$ and $\Phis$, respectively (note that for simplicity we neglected the possibility of any $\Delta$-dependence in the constants of proportionality). As $C(\sigma_x^1,\sigma_x^2)$ and/or $C(\sigma_x^1,\sigma_y^2)$ are non-zero, we can infer that $\Phi_c$ and/or $\Phi_s$ of the phase distribution are nonzero, which implies the ions are synchronized. This prediction is confirmed by $S$, plotted with the same parameters in Fig.~\ref{fig:pg4}(d), showing that synchronization can indeed be inferred using measurements of the spins alone (provided $\Phi_c$ or $\Phi_s$ are non-zero).

{\emph{Experimental Realization}.} Detecting correlations on the order of $10^{-3}$  is challenging, but possible with current technology. Projective
measurements of the ions' internal states \cite{Leibfried:2003,Blatt:2012} have been used to estimate observables with a precision over an order of magnitude higher than our requirement \cite{Harty:2014}. We now describe how the set-up shown in Fig.\ \ref{fig:pg1} and the parameter values used in our analysis could be achieved in practice. For concreteness we consider values of the Lamb-Dicke parameter $\eta=1/30$ and trap frequencies $(\omega_1+\omega_2)/2\gamma=500$ to ensure that the approximations used to derive Eq.\ (\ref{eqn:mainmast}) are valid. We have checked this explicitly by also calculating the correlation functions for these parameters using a version of Eq.\ (\ref{eqn:exactme}) in which we neglect only terms $O(\eta^4)$ and higher. The results are shown in Fig.\ \ref{fig:pg4} and overlay those obtained using Eq.\ (\ref{eqn:mainmast}).

The parameters we require could be achieved with two Ca$^+$ ions. Lasers at different wavelengths are readily available for manipulating their ground state $|4S\rangle$ and weakly excited states ($|4P_J\rangle$ and $|3D_J\rangle$ with $J$ the total angular momentum quantum number). Both the microtrap and the laser cooling technique have been demonstrated experimentally with this ion. The spin-down state is formed by the groundstate $|4S\rangle$ and spin-up state by the $|3D_{5/2}\rangle$ state, which are coupled by a standing-wave laser with a $729$ nm wavelength. This requires $\omega_j \approx 2\pi\times8.4$ MHz to have $\eta=1/30$. Given $\omega_j$, other quantities can be calculated straightforwardly. For example, the phonon coupling $J=2\pi\times1.68$ kHz, which is in the range that can be achieved in  current experiments~\cite{brown_2011,harlander_2011}.

We now explain how $\gamma$ can be tuned to the required value. For $(\omega_1+\omega_2)/2\gamma=500$, we should have $\gamma = 2\pi\times 16.8$ kHz. The natural decay rate of the $|3D_{5/2}\rangle$ state is about 1 Hz, which is far smaller than $\gamma$. In the following we show how the desired decay rate can be obtained by applying a quantum state engineering procedure~\cite{Genway2014}. We couple the $|3D_{5/2}\rangle$ state to the short-lived $|4P_{3/2}\rangle$ state by a dressing laser (854 nm). This laser dressing scheme allows us to tune the decay rate of the dressed state as $\gamma=[\Omega_D^2(\Gamma_1+\Gamma_2)]/[(\Gamma_1+\Gamma_2)^2+4\Delta_D^2]$. Here $\Delta_D$ and $\Omega_D$ are the detuning and Rabi frequency of the dressing laser. $\Gamma_1 =2\pi\times 135.1$ MHz and $\Gamma_2=2\pi\times9.9$ MHz are the decay rate from the $|4P_{3/2}\rangle$ state to the groundstate and to the $|3D_{5/2}\rangle$ state~\cite{Genway2014}, respectively. Using $\Delta_d=42$ MHz and $\Omega_d=2\pi\times1.8$ MHz, we finally have $\gamma=2\pi\times16.8$ kHz.

The phonon damping is realized by laser cooling techniques. Here we consider that the ions are cooled by a standing-wave laser. In this situation, the cooling rate depends on the position of the ions in the standing-wave pattern. For simplicity, we assume that the ions are located at nodes of the standing-wave~\cite{cirac1992}. The corresponding cooling rate is given by $\Gamma=\eta^2\Gamma_c[P(\Delta_c+\omega_j)-P(\Delta_c-\omega_j)]$, with $P(x)=\Omega_c^2/[4x^2+4\Gamma_c^2]$. Here $\Omega_c$ and $\Delta_c$ are the Rabi frequency and detuning of the cooling laser, and $\Gamma_c$ is the decay rate of the electronically excited state that is used in the laser cooling. We assume that the cooling laser couples the groundstate and the $|4P_{1/2}\rangle$ state, whose decay rate is $\Gamma_c\approx 2\pi\times 129.9$ MHz. Choosing $\Omega_c=1.0\,\Gamma_c$ and $\Delta_c=-2\pi\times100$ MHz, this leads to the required cooling rate $\Gamma\approx 2\pi\times5.6$ kHz.  Furthermore, these parameters allow us to adiabatically eliminate the state $|4P_{1/2}\rangle$ from coupling to the spins, as its dynamics happens on a much faster time scale than all the other processes. In addition, we have verified that the small difference between the two trapping frequencies will not change the cooling rate significantly. For example, the cooling rate of the two ions only differs by about $0.6\%$ when we look at the maximal difference between the trapping frequencies ($\Delta=3\gamma$), which can easily be compensated for by tuning the cooling laser parameters of individual ions.

\emph{Conclusions and Outlook.} We have shown that two phonon-lasing ions undergo synchronization when they are weakly coupled, leading to a bimodal relative phase distribution. Strong correlations develop between the internal, spin, degrees of freedom and the phonons in each ion and when the ions are coupled their synchronization leads to characteristic correlations \emph{between} the spins. These correlations carry information about the relative phase distribution of the ions and could be used infer the presence of synchronization.

The coupled ion phonon-laser system we consider is a promising model for future studies into quantum synchronization. Correlations between the spins of two ions at different times could be used to probe the dynamics of the synchronization process. Furthermore, the question of whether the non-classical (number-squeezed) phonon states that occur in this system might affect the development of synchronization, leading to significant differences compared to a corresponding semi-classical description\,\cite{Lee2013}, remains to be explored. It would also be interesting to compare the synchronization of ions forming a long chain in the quantum and classical regimes\,\cite{lee2011}.

\begin{acknowledgments}
This work was supported by the Leverhulme Trust (Grant No. F/00114/BG), EPSRC (Grant No. EP/I017828/1) and the European Research Council under the European Union's Seventh Framework Programme (FP/2007-2013) through ERC Grant Agreement No. 335266 (ESCQUMA) and the ERA-NET CHIST-ERA (R-ION consortium). We thank Matteo Marcuzzi, Katarzyna Macieszczak and the members of the R-ION consortium for insightful remarks. W.L. is supported through the Nottingham Research Fellowship by the University of Nottingham. MRH acknowledges funding from an Australian Research Council Discovery Project (project number DP110102322).
\end{acknowledgments}

\bibliographystyle{apsrev4-1}
\bibliography{ionsyncReferences}

\begin{thebibliography}{50}%
\makeatletter
\providecommand \@ifxundefined [1]{%
 \@ifx{#1\undefined}
}%
\providecommand \@ifnum [1]{%
 \ifnum #1\expandafter \@firstoftwo
 \else \expandafter \@secondoftwo
 \fi
}%
\providecommand \@ifx [1]{%
 \ifx #1\expandafter \@firstoftwo
 \else \expandafter \@secondoftwo
 \fi
}%
\providecommand \natexlab [1]{#1}%
\providecommand \enquote  [1]{``#1''}%
\providecommand \bibnamefont  [1]{#1}%
\providecommand \bibfnamefont [1]{#1}%
\providecommand \citenamefont [1]{#1}%
\providecommand \href@noop [0]{\@secondoftwo}%
\providecommand \href [0]{\begingroup \@sanitize@url \@href}%
\providecommand \@href[1]{\@@startlink{#1}\@@href}%
\providecommand \@@href[1]{\endgroup#1\@@endlink}%
\providecommand \@sanitize@url [0]{\catcode `\\12\catcode `\$12\catcode
  `\&12\catcode `\#12\catcode `\^12\catcode `\_12\catcode `\%12\relax}%
\providecommand \@@startlink[1]{}%
\providecommand \@@endlink[0]{}%
\providecommand \url  [0]{\begingroup\@sanitize@url \@url }%
\providecommand \@url [1]{\endgroup\@href {#1}{\urlprefix }}%
\providecommand \urlprefix  [0]{URL }%
\providecommand \Eprint [0]{\href }%
\providecommand \doibase [0]{http://dx.doi.org/}%
\providecommand \selectlanguage [0]{\@gobble}%
\providecommand \bibinfo  [0]{\@secondoftwo}%
\providecommand \bibfield  [0]{\@secondoftwo}%
\providecommand \translation [1]{[#1]}%
\providecommand \BibitemOpen [0]{}%
\providecommand \bibitemStop [0]{}%
\providecommand \bibitemNoStop [0]{.\EOS\space}%
\providecommand \EOS [0]{\spacefactor3000\relax}%
\providecommand \BibitemShut  [1]{\csname bibitem#1\endcsname}%
\let\auto@bib@innerbib\@empty
\bibitem [{\citenamefont {Pikovsky}\ \emph {et~al.}(2003)\citenamefont
  {Pikovsky}, \citenamefont {Rosenblum},\ and\ \citenamefont
  {Kurths}}]{Pikovsky2003}%
  \BibitemOpen
  \bibfield  {author} {\bibinfo {author} {\bibfnamefont {A.}~\bibnamefont
  {Pikovsky}}, \bibinfo {author} {\bibfnamefont {M.}~\bibnamefont {Rosenblum}},
  \ and\ \bibinfo {author} {\bibfnamefont {J.}~\bibnamefont {Kurths}},\ }\href
  {http://books.google.co.uk/books?id=FuIv845q3QUC} {\emph {\bibinfo {title}
  {Synchronization: A Universal Concept in Nonlinear Sciences}}},\ Cambridge
  Nonlinear Science Series\ (\bibinfo  {publisher} {Cambridge University
  Press},\ \bibinfo {year} {2003})\BibitemShut {NoStop}%
\bibitem [{\citenamefont {Roy}\ and\ \citenamefont
  {Thornburg}(1994)}]{Roy:1994}%
  \BibitemOpen
  \bibfield  {author} {\bibinfo {author} {\bibfnamefont {R.}~\bibnamefont
  {Roy}}\ and\ \bibinfo {author} {\bibfnamefont {K.~S.}\ \bibnamefont
  {Thornburg}},\ }\href {\doibase 10.1103/PhysRevLett.72.2009} {\bibfield
  {journal} {\bibinfo  {journal} {Phys. Rev. Lett.}\ }\textbf {\bibinfo
  {volume} {72}},\ \bibinfo {pages} {2009} (\bibinfo {year}
  {1994})}\BibitemShut {NoStop}%
\bibitem [{\citenamefont {DeShazer}\ \emph {et~al.}(2001)\citenamefont
  {DeShazer}, \citenamefont {Breban}, \citenamefont {Ott},\ and\ \citenamefont
  {Roy}}]{DeShazer:2001}%
  \BibitemOpen
  \bibfield  {author} {\bibinfo {author} {\bibfnamefont {D.~J.}\ \bibnamefont
  {DeShazer}}, \bibinfo {author} {\bibfnamefont {R.}~\bibnamefont {Breban}},
  \bibinfo {author} {\bibfnamefont {E.}~\bibnamefont {Ott}}, \ and\ \bibinfo
  {author} {\bibfnamefont {R.}~\bibnamefont {Roy}},\ }\href {\doibase
  10.1103/PhysRevLett.87.044101} {\bibfield  {journal} {\bibinfo  {journal}
  {Phys. Rev. Lett.}\ }\textbf {\bibinfo {volume} {87}},\ \bibinfo {pages}
  {044101} (\bibinfo {year} {2001})}\BibitemShut {NoStop}%
\bibitem [{\citenamefont {Heil}\ \emph {et~al.}(2001)\citenamefont {Heil},
  \citenamefont {Fischer}, \citenamefont {Els\"asser}, \citenamefont {Mulet},\
  and\ \citenamefont {Mirasso}}]{Heil:2001}%
  \BibitemOpen
  \bibfield  {author} {\bibinfo {author} {\bibfnamefont {T.}~\bibnamefont
  {Heil}}, \bibinfo {author} {\bibfnamefont {I.}~\bibnamefont {Fischer}},
  \bibinfo {author} {\bibfnamefont {W.}~\bibnamefont {Els\"asser}}, \bibinfo
  {author} {\bibfnamefont {J.}~\bibnamefont {Mulet}}, \ and\ \bibinfo {author}
  {\bibfnamefont {C.~R.}\ \bibnamefont {Mirasso}},\ }\href {\doibase
  10.1103/PhysRevLett.86.795} {\bibfield  {journal} {\bibinfo  {journal} {Phys.
  Rev. Lett.}\ }\textbf {\bibinfo {volume} {86}},\ \bibinfo {pages} {795}
  (\bibinfo {year} {2001})}\BibitemShut {NoStop}%
\bibitem [{\citenamefont {Wiesenfeld}\ \emph {et~al.}(1996)\citenamefont
  {Wiesenfeld}, \citenamefont {Colet},\ and\ \citenamefont
  {Strogatz}}]{Wiesenfeld:1996}%
  \BibitemOpen
  \bibfield  {author} {\bibinfo {author} {\bibfnamefont {K.}~\bibnamefont
  {Wiesenfeld}}, \bibinfo {author} {\bibfnamefont {P.}~\bibnamefont {Colet}}, \
  and\ \bibinfo {author} {\bibfnamefont {S.~H.}\ \bibnamefont {Strogatz}},\
  }\href {\doibase 10.1103/PhysRevLett.76.404} {\bibfield  {journal} {\bibinfo
  {journal} {Phys. Rev. Lett.}\ }\textbf {\bibinfo {volume} {76}},\ \bibinfo
  {pages} {404} (\bibinfo {year} {1996})}\BibitemShut {NoStop}%
\bibitem [{\citenamefont {Acebr\'on}\ \emph {et~al.}(2005)\citenamefont
  {Acebr\'on}, \citenamefont {Bonilla}, \citenamefont {P\'erez~Vicente},
  \citenamefont {Ritort},\ and\ \citenamefont {Spigler}}]{Acebron:2005}%
  \BibitemOpen
  \bibfield  {author} {\bibinfo {author} {\bibfnamefont {J.~A.}\ \bibnamefont
  {Acebr\'on}}, \bibinfo {author} {\bibfnamefont {L.~L.}\ \bibnamefont
  {Bonilla}}, \bibinfo {author} {\bibfnamefont {C.~J.}\ \bibnamefont
  {P\'erez~Vicente}}, \bibinfo {author} {\bibfnamefont {F.}~\bibnamefont
  {Ritort}}, \ and\ \bibinfo {author} {\bibfnamefont {R.}~\bibnamefont
  {Spigler}},\ }\href {\doibase 10.1103/RevModPhys.77.137} {\bibfield
  {journal} {\bibinfo  {journal} {Rev. Mod. Phys.}\ }\textbf {\bibinfo {volume}
  {77}},\ \bibinfo {pages} {137} (\bibinfo {year} {2005})}\BibitemShut
  {NoStop}%
\bibitem [{\citenamefont {Vinokur}\ \emph {et~al.}(2008)\citenamefont
  {Vinokur}, \citenamefont {Baturina}, \citenamefont {Fistul}, \citenamefont
  {Mironov}, \citenamefont {Baklanov},\ and\ \citenamefont
  {Strunk}}]{Vinokur:2008}%
  \BibitemOpen
  \bibfield  {author} {\bibinfo {author} {\bibfnamefont {V.~M.}\ \bibnamefont
  {Vinokur}}, \bibinfo {author} {\bibfnamefont {T.~I.}\ \bibnamefont
  {Baturina}}, \bibinfo {author} {\bibfnamefont {M.~V.}\ \bibnamefont
  {Fistul}}, \bibinfo {author} {\bibfnamefont {A.~Y.}\ \bibnamefont {Mironov}},
  \bibinfo {author} {\bibfnamefont {M.~R.}\ \bibnamefont {Baklanov}}, \ and\
  \bibinfo {author} {\bibfnamefont {C.}~\bibnamefont {Strunk}},\ }\href
  {http://dx.doi.org/10.1038/nature06837} {\bibfield  {journal} {\bibinfo
  {journal} {Nature}\ }\textbf {\bibinfo {volume} {452}},\ \bibinfo {pages}
  {613} (\bibinfo {year} {2008})}\BibitemShut {NoStop}%
\bibitem [{\citenamefont {Agrawal}\ \emph {et~al.}(2013)\citenamefont
  {Agrawal}, \citenamefont {Woodhouse},\ and\ \citenamefont
  {Seshia}}]{Agrawal:2013}%
  \BibitemOpen
  \bibfield  {author} {\bibinfo {author} {\bibfnamefont {D.~K.}\ \bibnamefont
  {Agrawal}}, \bibinfo {author} {\bibfnamefont {J.}~\bibnamefont {Woodhouse}},
  \ and\ \bibinfo {author} {\bibfnamefont {A.~A.}\ \bibnamefont {Seshia}},\
  }\href {\doibase 10.1103/PhysRevLett.111.084101} {\bibfield  {journal}
  {\bibinfo  {journal} {Phys. Rev. Lett.}\ }\textbf {\bibinfo {volume} {111}},\
  \bibinfo {pages} {084101} (\bibinfo {year} {2013})}\BibitemShut {NoStop}%
\bibitem [{\citenamefont {Matheny}\ \emph {et~al.}(2014)\citenamefont
  {Matheny}, \citenamefont {Grau}, \citenamefont {Villanueva}, \citenamefont
  {Karabalin}, \citenamefont {Cross},\ and\ \citenamefont
  {Roukes}}]{Roukes:2014}%
  \BibitemOpen
  \bibfield  {author} {\bibinfo {author} {\bibfnamefont {M.~H.}\ \bibnamefont
  {Matheny}}, \bibinfo {author} {\bibfnamefont {M.}~\bibnamefont {Grau}},
  \bibinfo {author} {\bibfnamefont {L.~G.}\ \bibnamefont {Villanueva}},
  \bibinfo {author} {\bibfnamefont {R.~B.}\ \bibnamefont {Karabalin}}, \bibinfo
  {author} {\bibfnamefont {M.~C.}\ \bibnamefont {Cross}}, \ and\ \bibinfo
  {author} {\bibfnamefont {M.~L.}\ \bibnamefont {Roukes}},\ }\href {\doibase
  10.1103/PhysRevLett.112.014101} {\bibfield  {journal} {\bibinfo  {journal}
  {Phys. Rev. Lett.}\ }\textbf {\bibinfo {volume} {112}},\ \bibinfo {pages}
  {014101} (\bibinfo {year} {2014})}\BibitemShut {NoStop}%
\bibitem [{\citenamefont {Gieseler}\ \emph {et~al.}(2014)\citenamefont
  {Gieseler}, \citenamefont {Spasenovi\ifmmode~\acute{c}\else \'{c}\fi{}},
  \citenamefont {Novotny},\ and\ \citenamefont {Quidant}}]{Giesler:2014}%
  \BibitemOpen
  \bibfield  {author} {\bibinfo {author} {\bibfnamefont {J.}~\bibnamefont
  {Gieseler}}, \bibinfo {author} {\bibfnamefont {M.}~\bibnamefont
  {Spasenovi\ifmmode~\acute{c}\else \'{c}\fi{}}}, \bibinfo {author}
  {\bibfnamefont {L.}~\bibnamefont {Novotny}}, \ and\ \bibinfo {author}
  {\bibfnamefont {R.}~\bibnamefont {Quidant}},\ }\href {\doibase
  10.1103/PhysRevLett.112.103603} {\bibfield  {journal} {\bibinfo  {journal}
  {Phys. Rev. Lett.}\ }\textbf {\bibinfo {volume} {112}},\ \bibinfo {pages}
  {103603} (\bibinfo {year} {2014})}\BibitemShut {NoStop}%
\bibitem [{\citenamefont {Lee}\ and\ \citenamefont
  {Sadeghpour}(2013)}]{Lee2013}%
  \BibitemOpen
  \bibfield  {author} {\bibinfo {author} {\bibfnamefont {T.~E.}\ \bibnamefont
  {Lee}}\ and\ \bibinfo {author} {\bibfnamefont {H.~R.}\ \bibnamefont
  {Sadeghpour}},\ }\href {\doibase 10.1103/PhysRevLett.111.234101} {\bibfield
  {journal} {\bibinfo  {journal} {Phys. Rev. Lett.}\ }\textbf {\bibinfo
  {volume} {111}},\ \bibinfo {pages} {234101} (\bibinfo {year}
  {2013})}\BibitemShut {NoStop}%
\bibitem [{\citenamefont {Lee}\ \emph {et~al.}(2014)\citenamefont {Lee},
  \citenamefont {Chan},\ and\ \citenamefont {Wang}}]{Lee2014}%
  \BibitemOpen
  \bibfield  {author} {\bibinfo {author} {\bibfnamefont {T.~E.}\ \bibnamefont
  {Lee}}, \bibinfo {author} {\bibfnamefont {C.-K.}\ \bibnamefont {Chan}}, \
  and\ \bibinfo {author} {\bibfnamefont {S.}~\bibnamefont {Wang}},\ }\href
  {\doibase 10.1103/PhysRevE.89.022913} {\bibfield  {journal} {\bibinfo
  {journal} {Phys. Rev. E}\ }\textbf {\bibinfo {volume} {89}},\ \bibinfo
  {pages} {022913} (\bibinfo {year} {2014})}\BibitemShut {NoStop}%
\bibitem [{\citenamefont {Walter}\ \emph {et~al.}(2015)\citenamefont {Walter},
  \citenamefont {Nunnenkamp},\ and\ \citenamefont {Bruder}}]{Walter:2014}%
  \BibitemOpen
  \bibfield  {author} {\bibinfo {author} {\bibfnamefont {S.}~\bibnamefont
  {Walter}}, \bibinfo {author} {\bibfnamefont {A.}~\bibnamefont {Nunnenkamp}},
  \ and\ \bibinfo {author} {\bibfnamefont {C.}~\bibnamefont {Bruder}},\ }\href
  {http://dx.doi.org/10.1002/andp.201400144} {\bibfield  {journal} {\bibinfo
  {journal} {Ann. Phys.}\ }\textbf {\bibinfo {volume} {527}},\ \bibinfo {pages}
  {131} (\bibinfo {year} {2015})}\BibitemShut {NoStop}%
\bibitem [{\citenamefont {Walter}\ \emph {et~al.}(2014)\citenamefont {Walter},
  \citenamefont {Nunnenkamp},\ and\ \citenamefont {Bruder}}]{Walter:2014a}%
  \BibitemOpen
  \bibfield  {author} {\bibinfo {author} {\bibfnamefont {S.}~\bibnamefont
  {Walter}}, \bibinfo {author} {\bibfnamefont {A.}~\bibnamefont {Nunnenkamp}},
  \ and\ \bibinfo {author} {\bibfnamefont {C.}~\bibnamefont {Bruder}},\ }\href
  {\doibase 10.1103/PhysRevLett.112.094102} {\bibfield  {journal} {\bibinfo
  {journal} {Phys. Rev. Lett.}\ }\textbf {\bibinfo {volume} {112}},\ \bibinfo
  {pages} {094102} (\bibinfo {year} {2014})}\BibitemShut {NoStop}%
\bibitem [{\citenamefont {Choi}\ and\ \citenamefont {Ha}(2014)}]{Choi:2014}%
  \BibitemOpen
  \bibfield  {author} {\bibinfo {author} {\bibfnamefont {S.-H.}\ \bibnamefont
  {Choi}}\ and\ \bibinfo {author} {\bibfnamefont {S.-Y.}\ \bibnamefont {Ha}},\
  }\href {http://stacks.iop.org/1751-8121/47/i=35/a=355104} {\bibfield
  {journal} {\bibinfo  {journal} {Journal of Physics A: Mathematical and
  Theoretical}\ }\textbf {\bibinfo {volume} {47}},\ \bibinfo {pages} {355104}
  (\bibinfo {year} {2014})}\BibitemShut {NoStop}%
\bibitem [{\citenamefont {Mari}\ \emph {et~al.}(2013)\citenamefont {Mari},
  \citenamefont {Farace}, \citenamefont {Didier}, \citenamefont {Giovannetti},\
  and\ \citenamefont {Fazio}}]{Mari2013}%
  \BibitemOpen
  \bibfield  {author} {\bibinfo {author} {\bibfnamefont {A.}~\bibnamefont
  {Mari}}, \bibinfo {author} {\bibfnamefont {A.}~\bibnamefont {Farace}},
  \bibinfo {author} {\bibfnamefont {N.}~\bibnamefont {Didier}}, \bibinfo
  {author} {\bibfnamefont {V.}~\bibnamefont {Giovannetti}}, \ and\ \bibinfo
  {author} {\bibfnamefont {R.}~\bibnamefont {Fazio}},\ }\href {\doibase
  10.1103/PhysRevLett.111.103605} {\bibfield  {journal} {\bibinfo  {journal}
  {Phys. Rev. Lett.}\ }\textbf {\bibinfo {volume} {111}},\ \bibinfo {pages}
  {103605} (\bibinfo {year} {2013})}\BibitemShut {NoStop}%
\bibitem [{\citenamefont {Xu}\ \emph {et~al.}(2014)\citenamefont {Xu},
  \citenamefont {Tieri}, \citenamefont {Fine}, \citenamefont {Thompson},\ and\
  \citenamefont {Holland}}]{Xu:2014}%
  \BibitemOpen
  \bibfield  {author} {\bibinfo {author} {\bibfnamefont {M.}~\bibnamefont
  {Xu}}, \bibinfo {author} {\bibfnamefont {D.~A.}\ \bibnamefont {Tieri}},
  \bibinfo {author} {\bibfnamefont {E.~C.}\ \bibnamefont {Fine}}, \bibinfo
  {author} {\bibfnamefont {J.~K.}\ \bibnamefont {Thompson}}, \ and\ \bibinfo
  {author} {\bibfnamefont {M.~J.}\ \bibnamefont {Holland}},\ }\href {\doibase
  10.1103/PhysRevLett.113.154101} {\bibfield  {journal} {\bibinfo  {journal}
  {Phys. Rev. Lett.}\ }\textbf {\bibinfo {volume} {113}},\ \bibinfo {pages}
  {154101} (\bibinfo {year} {2014})}\BibitemShut {NoStop}%
\bibitem [{\citenamefont {Ludwig}\ and\ \citenamefont
  {Marquardt}(2013)}]{Ludwig:2013}%
  \BibitemOpen
  \bibfield  {author} {\bibinfo {author} {\bibfnamefont {M.}~\bibnamefont
  {Ludwig}}\ and\ \bibinfo {author} {\bibfnamefont {F.}~\bibnamefont
  {Marquardt}},\ }\href {\doibase 10.1103/PhysRevLett.111.073603} {\bibfield
  {journal} {\bibinfo  {journal} {Phys. Rev. Lett.}\ }\textbf {\bibinfo
  {volume} {111}},\ \bibinfo {pages} {073603} (\bibinfo {year}
  {2013})}\BibitemShut {NoStop}%
\bibitem [{\citenamefont {Manzano}\ \emph {et~al.}(2013)\citenamefont
  {Manzano}, \citenamefont {Galve}, \citenamefont {Giorgi}, \citenamefont
  {Hern{\'a}ndez-Garc{\'\i}a},\ and\ \citenamefont {Zambrini}}]{Manzano2013}%
  \BibitemOpen
  \bibfield  {author} {\bibinfo {author} {\bibfnamefont {G.}~\bibnamefont
  {Manzano}}, \bibinfo {author} {\bibfnamefont {F.}~\bibnamefont {Galve}},
  \bibinfo {author} {\bibfnamefont {G.~L.}\ \bibnamefont {Giorgi}}, \bibinfo
  {author} {\bibfnamefont {E.}~\bibnamefont {Hern{\'a}ndez-Garc{\'\i}a}}, \
  and\ \bibinfo {author} {\bibfnamefont {R.}~\bibnamefont {Zambrini}},\ }\href
  {\doibase 10.1038/srep01439} {\bibfield  {journal} {\bibinfo  {journal}
  {Scientific Reports}\ }\textbf {\bibinfo {volume} {3}},\ \bibinfo {pages}
  {1439} (\bibinfo {year} {2013})}\BibitemShut {NoStop}%
\bibitem [{\citenamefont {{Hermoso de Mendoza}}\ \emph
  {et~al.}(2013)\citenamefont {{Hermoso de Mendoza}}, \citenamefont
  {{Pach{\'o}n}}, \citenamefont {{G{\'o}mez-Garde{\~n}es}},\ and\ \citenamefont
  {{Zueco}}}]{Mendoza2013}%
  \BibitemOpen
  \bibfield  {author} {\bibinfo {author} {\bibfnamefont {I.}~\bibnamefont
  {{Hermoso de Mendoza}}}, \bibinfo {author} {\bibfnamefont {L.~A.}\
  \bibnamefont {{Pach{\'o}n}}}, \bibinfo {author} {\bibfnamefont
  {J.}~\bibnamefont {{G{\'o}mez-Garde{\~n}es}}}, \ and\ \bibinfo {author}
  {\bibfnamefont {D.}~\bibnamefont {{Zueco}}},\ }\href@noop {} {\bibfield
  {journal} {\bibinfo  {journal} {ArXiv e-prints}\ } (\bibinfo {year}
  {2013})},\ \Eprint {http://arxiv.org/abs/1309.3972} {arXiv:1309.3972
  [cond-mat.stat-mech]} \BibitemShut {NoStop}%
\bibitem [{\citenamefont {Vahala}\ \emph {et~al.}(2009)\citenamefont {Vahala},
  \citenamefont {Herrmann}, \citenamefont {Kn{\"u}nz}, \citenamefont
  {Batteiger}, \citenamefont {Saathoff}, \citenamefont {H{\"a}nsch},\ and\
  \citenamefont {Udem}}]{Vahala2009}%
  \BibitemOpen
  \bibfield  {author} {\bibinfo {author} {\bibfnamefont {K.}~\bibnamefont
  {Vahala}}, \bibinfo {author} {\bibfnamefont {M.}~\bibnamefont {Herrmann}},
  \bibinfo {author} {\bibfnamefont {S.}~\bibnamefont {Kn{\"u}nz}}, \bibinfo
  {author} {\bibfnamefont {V.}~\bibnamefont {Batteiger}}, \bibinfo {author}
  {\bibfnamefont {G.}~\bibnamefont {Saathoff}}, \bibinfo {author}
  {\bibfnamefont {T.}~\bibnamefont {H{\"a}nsch}}, \ and\ \bibinfo {author}
  {\bibfnamefont {T.}~\bibnamefont {Udem}},\ }\href@noop {} {\bibfield
  {journal} {\bibinfo  {journal} {Nature Physics}\ }\textbf {\bibinfo {volume}
  {5}},\ \bibinfo {pages} {682} (\bibinfo {year} {2009})}\BibitemShut {NoStop}%
\bibitem [{\citenamefont {Kn\"unz}\ \emph {et~al.}(2010)\citenamefont
  {Kn\"unz}, \citenamefont {Herrmann}, \citenamefont {Batteiger}, \citenamefont
  {Saathoff}, \citenamefont {H\"ansch}, \citenamefont {Vahala},\ and\
  \citenamefont {Udem}}]{Knunz:2010}%
  \BibitemOpen
  \bibfield  {author} {\bibinfo {author} {\bibfnamefont {S.}~\bibnamefont
  {Kn\"unz}}, \bibinfo {author} {\bibfnamefont {M.}~\bibnamefont {Herrmann}},
  \bibinfo {author} {\bibfnamefont {V.}~\bibnamefont {Batteiger}}, \bibinfo
  {author} {\bibfnamefont {G.}~\bibnamefont {Saathoff}}, \bibinfo {author}
  {\bibfnamefont {T.~W.}\ \bibnamefont {H\"ansch}}, \bibinfo {author}
  {\bibfnamefont {K.}~\bibnamefont {Vahala}}, \ and\ \bibinfo {author}
  {\bibfnamefont {T.}~\bibnamefont {Udem}},\ }\href {\doibase
  10.1103/PhysRevLett.105.013004} {\bibfield  {journal} {\bibinfo  {journal}
  {Phys. Rev. Lett.}\ }\textbf {\bibinfo {volume} {105}},\ \bibinfo {pages}
  {013004} (\bibinfo {year} {2010})}\BibitemShut {NoStop}%
\bibitem [{\citenamefont {Xie}\ \emph {et~al.}(2013)\citenamefont {Xie},
  \citenamefont {Wan}, \citenamefont {Wu}, \citenamefont {Zhou}, \citenamefont
  {Chen},\ and\ \citenamefont {Feng}}]{Xie:2013}%
  \BibitemOpen
  \bibfield  {author} {\bibinfo {author} {\bibfnamefont {Y.}~\bibnamefont
  {Xie}}, \bibinfo {author} {\bibfnamefont {W.}~\bibnamefont {Wan}}, \bibinfo
  {author} {\bibfnamefont {H.~Y.}\ \bibnamefont {Wu}}, \bibinfo {author}
  {\bibfnamefont {F.}~\bibnamefont {Zhou}}, \bibinfo {author} {\bibfnamefont
  {L.}~\bibnamefont {Chen}}, \ and\ \bibinfo {author} {\bibfnamefont
  {M.}~\bibnamefont {Feng}},\ }\href {\doibase 10.1103/PhysRevA.87.053402}
  {\bibfield  {journal} {\bibinfo  {journal} {Phys. Rev. A}\ }\textbf {\bibinfo
  {volume} {87}},\ \bibinfo {pages} {053402} (\bibinfo {year}
  {2013})}\BibitemShut {NoStop}%
\bibitem [{\citenamefont {Brown}\ \emph {et~al.}(2011)\citenamefont {Brown},
  \citenamefont {Ospelkaus}, \citenamefont {Colombe}, \citenamefont {Wilson},
  \citenamefont {Leibfried},\ and\ \citenamefont {Wineland}}]{brown_2011}%
  \BibitemOpen
  \bibfield  {author} {\bibinfo {author} {\bibfnamefont {K.~R.}\ \bibnamefont
  {Brown}}, \bibinfo {author} {\bibfnamefont {C.}~\bibnamefont {Ospelkaus}},
  \bibinfo {author} {\bibfnamefont {Y.}~\bibnamefont {Colombe}}, \bibinfo
  {author} {\bibfnamefont {A.~C.}\ \bibnamefont {Wilson}}, \bibinfo {author}
  {\bibfnamefont {D.}~\bibnamefont {Leibfried}}, \ and\ \bibinfo {author}
  {\bibfnamefont {D.~J.}\ \bibnamefont {Wineland}},\ }\href {\doibase
  10.1038/nature09721} {\bibfield  {journal} {\bibinfo  {journal} {Nature}\
  }\textbf {\bibinfo {volume} {471}},\ \bibinfo {pages} {196} (\bibinfo {year}
  {2011})}\BibitemShut {NoStop}%
\bibitem [{\citenamefont {Harlander}\ \emph {et~al.}(2011)\citenamefont
  {Harlander}, \citenamefont {Lechner}, \citenamefont {Brownnutt},
  \citenamefont {Blatt},\ and\ \citenamefont {H{\"a}nsel}}]{harlander_2011}%
  \BibitemOpen
  \bibfield  {author} {\bibinfo {author} {\bibfnamefont {M.}~\bibnamefont
  {Harlander}}, \bibinfo {author} {\bibfnamefont {R.}~\bibnamefont {Lechner}},
  \bibinfo {author} {\bibfnamefont {M.}~\bibnamefont {Brownnutt}}, \bibinfo
  {author} {\bibfnamefont {R.}~\bibnamefont {Blatt}}, \ and\ \bibinfo {author}
  {\bibfnamefont {W.}~\bibnamefont {H{\"a}nsel}},\ }\href {\doibase
  10.1038/nature09800} {\bibfield  {journal} {\bibinfo  {journal} {Nature}\
  }\textbf {\bibinfo {volume} {471}},\ \bibinfo {pages} {200} (\bibinfo {year}
  {2011})}\BibitemShut {NoStop}%
\bibitem [{\citenamefont {Cirac}\ \emph
  {et~al.}(1992{\natexlab{a}})\citenamefont {Cirac}, \citenamefont {Blatt},
  \citenamefont {Zoller},\ and\ \citenamefont {Phillips}}]{Cirac:1992}%
  \BibitemOpen
  \bibfield  {author} {\bibinfo {author} {\bibfnamefont {J.~I.}\ \bibnamefont
  {Cirac}}, \bibinfo {author} {\bibfnamefont {R.}~\bibnamefont {Blatt}},
  \bibinfo {author} {\bibfnamefont {P.}~\bibnamefont {Zoller}}, \ and\ \bibinfo
  {author} {\bibfnamefont {W.~D.}\ \bibnamefont {Phillips}},\ }\href {\doibase
  10.1103/PhysRevA.46.2668} {\bibfield  {journal} {\bibinfo  {journal} {Phys.
  Rev. A}\ }\textbf {\bibinfo {volume} {46}},\ \bibinfo {pages} {2668}
  (\bibinfo {year} {1992}{\natexlab{a}})}\BibitemShut {NoStop}%
\bibitem [{\citenamefont {Marzoli}\ \emph {et~al.}(1994)\citenamefont
  {Marzoli}, \citenamefont {Cirac}, \citenamefont {Blatt},\ and\ \citenamefont
  {Zoller}}]{Marzoli:1994}%
  \BibitemOpen
  \bibfield  {author} {\bibinfo {author} {\bibfnamefont {I.}~\bibnamefont
  {Marzoli}}, \bibinfo {author} {\bibfnamefont {J.~I.}\ \bibnamefont {Cirac}},
  \bibinfo {author} {\bibfnamefont {R.}~\bibnamefont {Blatt}}, \ and\ \bibinfo
  {author} {\bibfnamefont {P.}~\bibnamefont {Zoller}},\ }\href {\doibase
  10.1103/PhysRevA.49.2771} {\bibfield  {journal} {\bibinfo  {journal} {Phys.
  Rev. A}\ }\textbf {\bibinfo {volume} {49}},\ \bibinfo {pages} {2771}
  (\bibinfo {year} {1994})}\BibitemShut {NoStop}%
\bibitem [{\citenamefont {Genway}\ \emph {et~al.}(2014)\citenamefont {Genway},
  \citenamefont {Li}, \citenamefont {Ates}, \citenamefont {Lanyon},\ and\
  \citenamefont {Lesanovsky}}]{Genway2014}%
  \BibitemOpen
  \bibfield  {author} {\bibinfo {author} {\bibfnamefont {S.}~\bibnamefont
  {Genway}}, \bibinfo {author} {\bibfnamefont {W.}~\bibnamefont {Li}}, \bibinfo
  {author} {\bibfnamefont {C.}~\bibnamefont {Ates}}, \bibinfo {author}
  {\bibfnamefont {B.~P.}\ \bibnamefont {Lanyon}}, \ and\ \bibinfo {author}
  {\bibfnamefont {I.}~\bibnamefont {Lesanovsky}},\ }\href {\doibase
  10.1103/PhysRevLett.112.023603} {\bibfield  {journal} {\bibinfo  {journal}
  {Phys. Rev. Lett.}\ }\textbf {\bibinfo {volume} {112}},\ \bibinfo {pages}
  {023603} (\bibinfo {year} {2014})}\BibitemShut {NoStop}%
\bibitem [{\citenamefont {Johansson}\ \emph {et~al.}(2012)\citenamefont
  {Johansson}, \citenamefont {Nation},\ and\ \citenamefont
  {Nori}}]{Johansson:2012}%
  \BibitemOpen
  \bibfield  {author} {\bibinfo {author} {\bibfnamefont {J.}~\bibnamefont
  {Johansson}}, \bibinfo {author} {\bibfnamefont {P.}~\bibnamefont {Nation}}, \
  and\ \bibinfo {author} {\bibfnamefont {F.}~\bibnamefont {Nori}},\ }\href
  {\doibase http://dx.doi.org/10.1016/j.cpc.2012.02.021} {\bibfield  {journal}
  {\bibinfo  {journal} {Computer Physics Communications}\ }\textbf {\bibinfo
  {volume} {183}},\ \bibinfo {pages} {1760 } (\bibinfo {year}
  {2012})}\BibitemShut {NoStop}%
\bibitem [{\citenamefont {Johansson}\ \emph {et~al.}(2013)\citenamefont
  {Johansson}, \citenamefont {Nation},\ and\ \citenamefont
  {Nori}}]{Johansson:2013}%
  \BibitemOpen
  \bibfield  {author} {\bibinfo {author} {\bibfnamefont {J.}~\bibnamefont
  {Johansson}}, \bibinfo {author} {\bibfnamefont {P.}~\bibnamefont {Nation}}, \
  and\ \bibinfo {author} {\bibfnamefont {F.}~\bibnamefont {Nori}},\ }\href
  {\doibase http://dx.doi.org/10.1016/j.cpc.2012.11.019} {\bibfield  {journal}
  {\bibinfo  {journal} {Computer Physics Communications}\ }\textbf {\bibinfo
  {volume} {184}},\ \bibinfo {pages} {1234 } (\bibinfo {year}
  {2013})}\BibitemShut {NoStop}%
\bibitem [{\citenamefont {Gardiner}\ and\ \citenamefont
  {Zoller}(2004)}]{Gardiner:2004}%
  \BibitemOpen
  \bibfield  {author} {\bibinfo {author} {\bibfnamefont {C.~W.}\ \bibnamefont
  {Gardiner}}\ and\ \bibinfo {author} {\bibfnamefont {P.}~\bibnamefont
  {Zoller}},\ }\href@noop {} {\emph {\bibinfo {title} {Quantum Noise: A
  Handbook of Markovian and Non-Markovian Quantum Stochastic Methods with
  Application to Quantum Optics}}},\ \bibinfo {edition} {3rd}\ ed.\ (\bibinfo
  {publisher} {Springer},\ \bibinfo {year} {2004})\BibitemShut {NoStop}%
\bibitem [{\citenamefont {Dubin}\ \emph {et~al.}(2010)\citenamefont {Dubin},
  \citenamefont {Russo}, \citenamefont {Barros}, \citenamefont {Stute},
  \citenamefont {Becher}, \citenamefont {Schmidt},\ and\ \citenamefont
  {Blatt}}]{Dubin:2010}%
  \BibitemOpen
  \bibfield  {author} {\bibinfo {author} {\bibfnamefont {F.}~\bibnamefont
  {Dubin}}, \bibinfo {author} {\bibfnamefont {C.}~\bibnamefont {Russo}},
  \bibinfo {author} {\bibfnamefont {H.~G.}\ \bibnamefont {Barros}}, \bibinfo
  {author} {\bibfnamefont {A.}~\bibnamefont {Stute}}, \bibinfo {author}
  {\bibfnamefont {C.}~\bibnamefont {Becher}}, \bibinfo {author} {\bibfnamefont
  {P.~O.}\ \bibnamefont {Schmidt}}, \ and\ \bibinfo {author} {\bibfnamefont
  {R.}~\bibnamefont {Blatt}},\ }\href {http://dx.doi.org/10.1038/nphys1627}
  {\bibfield  {journal} {\bibinfo  {journal} {Nat Phys}\ }\textbf {\bibinfo
  {volume} {6}},\ \bibinfo {pages} {350} (\bibinfo {year} {2010})}\BibitemShut
  {NoStop}%
\bibitem [{\citenamefont {Husimi}(1940)}]{Husimi:1940}%
  \BibitemOpen
  \bibfield  {author} {\bibinfo {author} {\bibfnamefont {K.}~\bibnamefont
  {Husimi}},\ }\href@noop {} {\bibfield  {journal} {\bibinfo  {journal} {Proc.
  Phys. Math. Soc. Jpn. 22:264-314}\ } (\bibinfo {year} {1940})}\BibitemShut
  {NoStop}%
\bibitem [{\citenamefont {Sudarshan}(1963)}]{Sudarshan:1963}%
  \BibitemOpen
  \bibfield  {author} {\bibinfo {author} {\bibfnamefont {E.~C.~G.}\
  \bibnamefont {Sudarshan}},\ }\href {\doibase 10.1103/PhysRevLett.10.277}
  {\bibfield  {journal} {\bibinfo  {journal} {Phys. Rev. Lett.}\ }\textbf
  {\bibinfo {volume} {10}},\ \bibinfo {pages} {277} (\bibinfo {year}
  {1963})}\BibitemShut {NoStop}%
\bibitem [{\citenamefont {Drummond}\ and\ \citenamefont
  {Gardiner}(1980)}]{Drummond:1980}%
  \BibitemOpen
  \bibfield  {author} {\bibinfo {author} {\bibfnamefont {P.~D.}\ \bibnamefont
  {Drummond}}\ and\ \bibinfo {author} {\bibfnamefont {C.~W.}\ \bibnamefont
  {Gardiner}},\ }\href {http://stacks.iop.org/0305-4470/13/i=7/a=018}
  {\bibfield  {journal} {\bibinfo  {journal} {Journal of Physics A:
  Mathematical and General}\ }\textbf {\bibinfo {volume} {13}},\ \bibinfo
  {pages} {2353} (\bibinfo {year} {1980})}\BibitemShut {NoStop}%
\bibitem [{\citenamefont {Vaccaro}(1995)}]{Vaccaro:1995}%
  \BibitemOpen
  \bibfield  {author} {\bibinfo {author} {\bibfnamefont {J.}~\bibnamefont
  {Vaccaro}},\ }\href {\doibase 10.1103/PhysRevA.52.3474} {\bibfield  {journal}
  {\bibinfo  {journal} {Phys. Rev. A}\ }\textbf {\bibinfo {volume} {52}},\
  \bibinfo {pages} {3474} (\bibinfo {year} {1995})}\BibitemShut {NoStop}%
\bibitem [{\citenamefont {Vaccaro}\ and\ \citenamefont
  {Pegg}(1990)}]{Vaccaro:1990}%
  \BibitemOpen
  \bibfield  {author} {\bibinfo {author} {\bibfnamefont {J.~A.}\ \bibnamefont
  {Vaccaro}}\ and\ \bibinfo {author} {\bibfnamefont {D.~T.}\ \bibnamefont
  {Pegg}},\ }\href {\doibase 10.1103/PhysRevA.41.5156} {\bibfield  {journal}
  {\bibinfo  {journal} {Phys. Rev. A}\ }\textbf {\bibinfo {volume} {41}},\
  \bibinfo {pages} {5156} (\bibinfo {year} {1990})}\BibitemShut {NoStop}%
\bibitem [{\citenamefont {Moya-Cessa}(2003)}]{Moya-Cessa:2003}%
  \BibitemOpen
  \bibfield  {author} {\bibinfo {author} {\bibfnamefont {H.}~\bibnamefont
  {Moya-Cessa}},\ }\href {http://stacks.iop.org/1464-4266/5/i=3/a=367}
  {\bibfield  {journal} {\bibinfo  {journal} {Journal of Optics B: Quantum and
  Semiclassical Optics}\ }\textbf {\bibinfo {volume} {5}},\ \bibinfo {pages}
  {S339} (\bibinfo {year} {2003})}\BibitemShut {NoStop}%
\bibitem [{\citenamefont {Hush}\ \emph {et~al.}(2010)\citenamefont {Hush},
  \citenamefont {Carvalho},\ and\ \citenamefont {Hope}}]{Hush:2010}%
  \BibitemOpen
  \bibfield  {author} {\bibinfo {author} {\bibfnamefont {M.~R.}\ \bibnamefont
  {Hush}}, \bibinfo {author} {\bibfnamefont {A.~R.~R.}\ \bibnamefont
  {Carvalho}}, \ and\ \bibinfo {author} {\bibfnamefont {J.~J.}\ \bibnamefont
  {Hope}},\ }\href {\doibase 10.1103/PhysRevA.81.033852} {\bibfield  {journal}
  {\bibinfo  {journal} {Phys. Rev. A}\ }\textbf {\bibinfo {volume} {81}},\
  \bibinfo {pages} {033852} (\bibinfo {year} {2010})}\BibitemShut {NoStop}%
\bibitem [{\citenamefont {Barak}\ and\ \citenamefont
  {Ben-Aryeh}(2005)}]{Barak:2005}%
  \BibitemOpen
  \bibfield  {author} {\bibinfo {author} {\bibfnamefont {R.}~\bibnamefont
  {Barak}}\ and\ \bibinfo {author} {\bibfnamefont {Y.}~\bibnamefont
  {Ben-Aryeh}},\ }\href {http://stacks.iop.org/1464-4266/7/i=5/a=001}
  {\bibfield  {journal} {\bibinfo  {journal} {Journal of Optics B: Quantum and
  Semiclassical Optics}\ }\textbf {\bibinfo {volume} {7}},\ \bibinfo {pages}
  {123} (\bibinfo {year} {2005})}\BibitemShut {NoStop}%
\bibitem [{\citenamefont {Barnett}\ and\ \citenamefont
  {Vaccaro}(2007)}]{QOp:2007}%
  \BibitemOpen
  \bibinfo {editor} {\bibfnamefont {S.~M.}\ \bibnamefont {Barnett}}\ and\
  \bibinfo {editor} {\bibfnamefont {J.~A.}\ \bibnamefont {Vaccaro}},\ eds.,\
  \href@noop {} {\emph {\bibinfo {title} {The Quantum Phase Operator: A
  Review}}}\ (\bibinfo  {publisher} {Taylor and Francis},\ \bibinfo {year}
  {2007})\BibitemShut {NoStop}%
\bibitem [{\citenamefont {Pegg}\ and\ \citenamefont
  {Barnett}(1988)}]{Pegg:1988}%
  \BibitemOpen
  \bibfield  {author} {\bibinfo {author} {\bibfnamefont {D.~T.}\ \bibnamefont
  {Pegg}}\ and\ \bibinfo {author} {\bibfnamefont {S.~M.}\ \bibnamefont
  {Barnett}},\ }\href {http://stacks.iop.org/0295-5075/6/i=6/a=002} {\bibfield
  {journal} {\bibinfo  {journal} {EPL (Europhysics Letters)}\ }\textbf
  {\bibinfo {volume} {6}},\ \bibinfo {pages} {483} (\bibinfo {year}
  {1988})}\BibitemShut {NoStop}%
\bibitem [{\citenamefont {Shapiro}\ and\ \citenamefont
  {Shepard}(1991)}]{Shapiro:1991}%
  \BibitemOpen
  \bibfield  {author} {\bibinfo {author} {\bibfnamefont {J.~H.}\ \bibnamefont
  {Shapiro}}\ and\ \bibinfo {author} {\bibfnamefont {S.~R.}\ \bibnamefont
  {Shepard}},\ }\href {\doibase 10.1103/PhysRevA.43.3795} {\bibfield  {journal}
  {\bibinfo  {journal} {Phys. Rev. A}\ }\textbf {\bibinfo {volume} {43}},\
  \bibinfo {pages} {3795} (\bibinfo {year} {1991})}\BibitemShut {NoStop}%
\bibitem [{\citenamefont {Kuznetsov}\ \emph {et~al.}(2009)\citenamefont
  {Kuznetsov}, \citenamefont {Stankevich},\ and\ \citenamefont
  {Turukina}}]{Kuznetsov:2009}%
  \BibitemOpen
  \bibfield  {author} {\bibinfo {author} {\bibfnamefont {A.}~\bibnamefont
  {Kuznetsov}}, \bibinfo {author} {\bibfnamefont {N.}~\bibnamefont
  {Stankevich}}, \ and\ \bibinfo {author} {\bibfnamefont {L.}~\bibnamefont
  {Turukina}},\ }\href {\doibase http://dx.doi.org/10.1016/j.physd.2009.04.001}
  {\bibfield  {journal} {\bibinfo  {journal} {Physica D: Nonlinear Phenomena}\
  }\textbf {\bibinfo {volume} {238}},\ \bibinfo {pages} {1203 } (\bibinfo
  {year} {2009})}\BibitemShut {NoStop}%
\bibitem [{\citenamefont {Ivanchenko}\ \emph {et~al.}(2004)\citenamefont
  {Ivanchenko}, \citenamefont {Osipov}, \citenamefont {Shalfeev},\ and\
  \citenamefont {Kurths}}]{Ivanchenko:2004}%
  \BibitemOpen
  \bibfield  {author} {\bibinfo {author} {\bibfnamefont {M.}~\bibnamefont
  {Ivanchenko}}, \bibinfo {author} {\bibfnamefont {G.}~\bibnamefont {Osipov}},
  \bibinfo {author} {\bibfnamefont {V.}~\bibnamefont {Shalfeev}}, \ and\
  \bibinfo {author} {\bibfnamefont {J.}~\bibnamefont {Kurths}},\ }\href
  {\doibase http://dx.doi.org/10.1016/j.physd.2003.09.035} {\bibfield
  {journal} {\bibinfo  {journal} {Physica D: Nonlinear Phenomena}\ }\textbf
  {\bibinfo {volume} {189}},\ \bibinfo {pages} {8 } (\bibinfo {year}
  {2004})}\BibitemShut {NoStop}%
\bibitem [{\citenamefont {Leibfried}\ \emph {et~al.}(2003)\citenamefont
  {Leibfried}, \citenamefont {Blatt}, \citenamefont {Monroe},\ and\
  \citenamefont {Wineland}}]{Leibfried:2003}%
  \BibitemOpen
  \bibfield  {author} {\bibinfo {author} {\bibfnamefont {D.}~\bibnamefont
  {Leibfried}}, \bibinfo {author} {\bibfnamefont {R.}~\bibnamefont {Blatt}},
  \bibinfo {author} {\bibfnamefont {C.}~\bibnamefont {Monroe}}, \ and\ \bibinfo
  {author} {\bibfnamefont {D.}~\bibnamefont {Wineland}},\ }\href {\doibase
  10.1103/RevModPhys.75.281} {\bibfield  {journal} {\bibinfo  {journal} {Rev.
  Mod. Phys.}\ }\textbf {\bibinfo {volume} {75}},\ \bibinfo {pages} {281}
  (\bibinfo {year} {2003})}\BibitemShut {NoStop}%
\bibitem [{\citenamefont {Blatt}\ and\ \citenamefont
  {Roos}(2012)}]{Blatt:2012}%
  \BibitemOpen
  \bibfield  {author} {\bibinfo {author} {\bibfnamefont {R.}~\bibnamefont
  {Blatt}}\ and\ \bibinfo {author} {\bibfnamefont {C.~F.}\ \bibnamefont
  {Roos}},\ }\href {http://dx.doi.org/10.1038/nphys2252} {\bibfield  {journal}
  {\bibinfo  {journal} {Nat Phys}\ }\textbf {\bibinfo {volume} {8}},\ \bibinfo
  {pages} {277} (\bibinfo {year} {2012})}\BibitemShut {NoStop}%
\bibitem [{\citenamefont {Harty}\ \emph {et~al.}(2014)\citenamefont {Harty},
  \citenamefont {Allcock}, \citenamefont {Ballance}, \citenamefont {Guidoni},
  \citenamefont {Janacek}, \citenamefont {Linke}, \citenamefont {Stacey},\ and\
  \citenamefont {Lucas}}]{Harty:2014}%
  \BibitemOpen
  \bibfield  {author} {\bibinfo {author} {\bibfnamefont {T.~P.}\ \bibnamefont
  {Harty}}, \bibinfo {author} {\bibfnamefont {D.~T.~C.}\ \bibnamefont
  {Allcock}}, \bibinfo {author} {\bibfnamefont {C.~J.}\ \bibnamefont
  {Ballance}}, \bibinfo {author} {\bibfnamefont {L.}~\bibnamefont {Guidoni}},
  \bibinfo {author} {\bibfnamefont {H.~A.}\ \bibnamefont {Janacek}}, \bibinfo
  {author} {\bibfnamefont {N.~M.}\ \bibnamefont {Linke}}, \bibinfo {author}
  {\bibfnamefont {D.~N.}\ \bibnamefont {Stacey}}, \ and\ \bibinfo {author}
  {\bibfnamefont {D.~M.}\ \bibnamefont {Lucas}},\ }\href@noop {} {\enquote
  {\bibinfo {title} {High-fidelity preparation, gates, memory and readout of a
  trapped-ion quantum bit},}\ } (\bibinfo {year} {2014}),\ \bibinfo {note}
  {arXiv:1403.1524}\BibitemShut {NoStop}%
\bibitem [{\citenamefont {Cirac}\ \emph
  {et~al.}(1992{\natexlab{b}})\citenamefont {Cirac}, \citenamefont {Blatt},
  \citenamefont {Zoller},\ and\ \citenamefont {Phillips}}]{cirac1992}%
  \BibitemOpen
  \bibfield  {author} {\bibinfo {author} {\bibfnamefont {J.~I.}\ \bibnamefont
  {Cirac}}, \bibinfo {author} {\bibfnamefont {R.}~\bibnamefont {Blatt}},
  \bibinfo {author} {\bibfnamefont {P.}~\bibnamefont {Zoller}}, \ and\ \bibinfo
  {author} {\bibfnamefont {W.~D.}\ \bibnamefont {Phillips}},\ }\href {\doibase
  10.1103/PhysRevA.46.2668} {\bibfield  {journal} {\bibinfo  {journal} {Phys.
  Rev. A}\ }\textbf {\bibinfo {volume} {46}},\ \bibinfo {pages} {2668}
  (\bibinfo {year} {1992}{\natexlab{b}})}\BibitemShut {NoStop}%
\bibitem [{\citenamefont {Lee}\ and\ \citenamefont {Cross}(2011)}]{lee2011}%
  \BibitemOpen
  \bibfield  {author} {\bibinfo {author} {\bibfnamefont {T.~E.}\ \bibnamefont
  {Lee}}\ and\ \bibinfo {author} {\bibfnamefont {M.~C.}\ \bibnamefont
  {Cross}},\ }\href {\doibase 10.1103/PhysRevLett.106.143001} {\bibfield
  {journal} {\bibinfo  {journal} {Phys. Rev. Lett.}\ }\textbf {\bibinfo
  {volume} {106}},\ \bibinfo {pages} {143001} (\bibinfo {year}
  {2011})}\BibitemShut {NoStop}%
\end{thebibliography}%

\end{document}